# Virtual monochromatic imaging in dual-source and dual-energy CT for visualization of acute ischemic stroke

**Hidetake Hara and Hiroshi Muraishi**

*School of Allied Health Sciences, Kitasato University, Japan 252-0373*

**Hiroki Matsuzawa**

*General Medical Center, Saitama Medical University, Japan 350-8550*

**Toshiyuki Inoue and Yasuo Nakajima**

*Department of Radiology, St. Marianna University School of Medicine, Japan 216-8511*

**Hitoshi Satoh and Shinji Abe**

*School of Health Sciences, Ibaraki Prefectural University of Health Sciences, Japan 300-0394*

We have recently developed a phantom that simulates acute ischemic stroke. We attempted to visualize acute-stage cerebral infarction by applying virtual monochromatic images to this phantom using dual-energy CT (DECT). Virtual monochromatic images were created using DECT from 40 to 100 keV at every 10 keV and from 60 to 80 keV at every 1 keV, under three energy conditions of tube voltages with thin (Sn) filters. Calculation of the CNR values allowed us to evaluate the visualization of acute-stage cerebral infarction. The CNR value of a virtual monochromatic image was the highest at 68 keV under 80 kV / Sn 140 kV, at 72 keV under 100 kV / Sn 140 kV, and at 67 keV under 140 kV / 80 kV. The CNR values of virtual monochromatic images between 65 and 75 keV were significantly higher than those obtained for all other created energy images. Therefore, optimal conditions for visualizing acute ischemic stroke were achievable.

Email: harah@kitasato-u.ac.jp

Fax: +81-42-778-9628

# I. INTRODUCTION

In Japan, cerebrovascular disease is the fourth cause of death and the first cause of bedridden patients. Recently, cerebral stroke has been under serious consideration due to the westernization of the diets and an increase in geriatric diseases. At present, cerebral stroke is still one of the most important diseases to control and prevent. To establish a method for image-based diagnosis of cerebral stroke, the authors developed a phantom that could correctly evaluate disease detection by image-processing and that could visualize disease using X-ray CT imaging, while evaluating the imaging conditions [1-4]. Visualization of acute cerebral stroke within 4.5 hours following the development of cerebral stroke is essential because it is the time index used to judge whether thrombolytic therapy with use of rt-PA is applicable [5]. Traditionally, it has been difficult to visualize acute cerebral stroke from images produced by X-ray CT [6]. This study attempted to improve the contrast of acute ischemic stroke using virtual monochromatic images from dual-energy CT in order to achieve accurate visualization within the limitation of 4.5 hours.

In general, the human body, in which multiple substances are mixed, is photographed using X-rays with multi-colored energy. Since the mass attenuation coefficient of the X-ray effective energy possessed by each substance cannot be correctly calculated, there are cases where different substances in the body possess the same CT value. Dual-Energy CT (see Fig.1) is one of the methods available to solve this phenomenon. Since the mass attenuation coefficient depends on X-ray energy, images with different CT values can be obtained at different tube voltages [2,7-9]. In this study, dual-energy CT images were obtained using different energy levels in X-ray CT imaging, and virtual monochromatic images were obtained using the dual-energy CT images. The study investigated the possibility of visualizing acute cerebral infarction using the virtual monochromatic images to achieve improvement in the contrast of the brain affected by the cerebral infarction.

## II. MATERIALS AND METHODS

Using a cerebral stroke phantom, which had been developed by the authors, an imitation disease with 32 HU (Hounsfield unit) was set to be visualized. The phantom, prepared with polyurethane resin and epoxy resin, was composed of three sections: brain, cranium, and imitation disease. The CT value of the brain section was 36 HU and that of the cranium was 900 HU (Table 1). The imitation disease section was composed of two balls, 20, 30 mm in diameter. The CT values of the two balls were 32, 34 HU were allocated in the middle cerebral artery region. The balls with 32 and 34 HU reflected acute cerebral stroke and their CT values were lower than that of the brain by 4 and 2 HU, respectively. (Fig. 2, 3) [1,3]. Using tube voltages at 80, 100, and 140 kV, which were assumed to be used clinically in X-ray CT imaging and a mAs value (tube current time product: mA-second) at 400, 600, 800 mAs, images with a slice thickness of 20 mm were obtained. Virtual monochromatic images were created using DECT from 40 to 100 keV at every 10 keV and from 60 to 80 keV at every 1 keV, under three energy conditions of tube voltages with thin (Sn) filters (i.e., 80 kV / Sn 140 kV, 100 kV / Sn 140 kV, and 140 kV / 80 kV). Because of this conditions used the signal-to-noise (S/N) ratio of the images of head region are generally reported to increase between 65 and 70 keV [10]. Calculation of the contrast-to-noise ratio (CNR) values allowed us to evaluate the visualization of acute-stage cerebral infarction. The CNR value at the 30-mm-diameter imitation disease section in each virtual monochromatic image was obtained using the formula ($| ROI_M - ROI_B |$) / $SD_B$, and the ability to visualize acute cerebral stroke was evaluated. In the formula, $ROI_M$ represents the CT value of the imitation disease section, $ROI_B$ represents the CT value of the brain section, and $SD_B$ represents the standard deviation (SD) value of the brain section [2,7]. In addition, an image with a CNR value exceeding 1.0 was defined as the ability to visualize acute cerebral stroke [1,4].

## III. RESULTS AND DISCUSSION

Figure 4, 5 and 6 show the simulation results of virtual monochromatic images. It's the cerebral infarction is clearly visible at 60 and 70 keV by 80 kV / Sn 140 kV form these created virtual monochromatic images, and the noise was generally reduced and the cerebral infarction is clearly visible at 60 and 70 keV by 100 kV / Sn 140 kV, and the contrast of the imitation disease section is markedly increased at 60 and 70 keV by 140 kV / 80 kV. Figure 7 and 8 show the simulation results of CNR values for each virtual monochromatic image. The CNR values of virtual monochromatic images in all tube voltage patterns were higher, reaching between 65 and 75 keV. The CNR value of a virtual monochromatic image was the highest at 68 keV under 80 kV / Sn 140 kV, at 72 keV under 100 kV / Sn 140 kV, and at 67 keV under 140 kV / 80 kV (see Fig.9). The CNR values of images with a slice thickness of 20 mm were greater than 1.0, which we defined as the ability to visualize acute-stage cerebral infarction by the authors [1-4]. Exceptions were the images obtained at 400 mAs under 80 kV / Sn 140 kV. In each tube voltage pattern, the CNR value of the virtual monochromatic image was higher than that of the original image. In this study, by using virtual monochromatic images, noise could be reduced while maintaining contrast at a specified level. The conditions required for virtual monochromatic imaging for visualization of acute-stage cerebral infarction are as follows: 1. Tube voltage combinations of 100 kV / Sn 140 kV and 140 kV / 80 kV, 2. A tube current time product of 600 mAs or higher and a slice thickness of 20 mm, 3. Virtual monochromatic imaging at approximately 70 keV.

## IV. CONCLUSION

We are able to detect acute ischemic stroke at its early stage by applying virtual monochromatic imaging to a phantom using dual-energy CT. The CNR values of virtual monochromatic images between 65 and 75 keV were significantly higher than those obtained for all other created energy images. Therefore, optimal conditions for visualizing acute-stage cerebral infarction were achievable. Dual-Energy CT images could be used clinically to judge whether thrombolytic therapy with the use of rt-PA is applicable to acute ischemic stroke.

## ACKNOWLEDGEMENT

The authors thank K. Imaizumi, A. Sano, R. Suzuki, M. Yoshida in Kitasato University for experimental support in this work. We also thank all the staff at General Medical Center, Saitama Medical University for their kind support of this work. This study was partially supported by grants the scientific research (C) from the Ministry of Education, Culture, Sports, Science and Technology of Japan (Grant-in-Aid No.26460732).

Table 1. Elemental compositions of the X-ray CT phantom to evaluate cerebral stroke.

|  | Spheres | | | Annulus |
| --- | --- | --- | --- | --- |
| Element | 36HU | 34HU | 32HU | 900HU |
| H | 8.16 | 8.16 | 8.17 | 7.74 |
| C | 70.00 | 70.08 | 70.15 | 67.31 |
| N | 4.47 | 4.47 | 4.48 | 2.74 |
| O | 15.56 | 15.53 | 15.50 | 18.10 |
| P | 0.58 | 0.56 | 0.54 | 0.0 |
| S | 0.0 | 0.0 | 0.0 | 0.78 |
| Ca | 1.24 | 1.20 | 1.16 | 0.0 |
| Ba | 0.0 | 0.0 | 0.0 | 3.33 |

Data are in wt %

Figure Captions.

Fig. 1. The externals of Dual-energy CT, by SOMATOM Definition Flash 128 DAS in Siemens Corporation.

Fig. 2. The plans of the X-ray CT phantom to evaluate cerebral stroke, which was prepared to have cranium and a composition and size similar to those of the brain section and acute cerebral stroke section.

Fig. 3. The externals of the X-ray CT phantom to evaluate cerebral stroke obtained from (a) lateral view, (b) frontal view.

Fig. 4. Virtual monochromatic images from 40 to 100 keV in 10 keV steps obtained at 80 kV / Sn 140 kV, (a) 40 keV, (b) 50 keV, (c) 60 keV, (d) 70 keV, (e) 80 keV, (f) 90 keV, (g) 100 keV. because of 20 mm in image slice thickness and 600 mAs.

Fig. 5. Virtual monochromatic images from 40 to 100 keV in 10 keV steps obtained at 100 kV / Sn 140 kV, (a) 40 keV, (b) 50 keV, (c) 60 keV, (d) 70 keV, (e) 80 keV, (f) 90 keV, (g) 100 keV. because of 20 mm in image slice thickness and 600 mAs.

Fig. 6. Virtual monochromatic images from 40 to 100 keV in 10 keV steps obtained at 140 kV / 80 kV, (a) 40 keV, (b) 50 keV, (c) 60 keV, (d) 70 keV, (e) 80 keV, (f) 90 keV, (g) 100 keV. because of 20 mm in image slice thickness and 600 mAs.

Fig. 7. Effect of virtual monochromatic images from 40 to 100 keV at every 10 keV in (a) 400 mAs, (b) 600 mAs, (c) 800 mAs. under three energy conditions of 80 kV / Sn 140 kV, 100 kV / Sn 140 kV, and 140 kV / 80 kV, on the contrast-to-noise ratio.

Fig. 8. Effect of virtual monochromatic images from 60 to 80 keV at every 1 keV in (a) 400 mAs, (b) 600 mAs, (c) 800 mAs. under three energy conditions of 80 kV / Sn 140 kV, 100 kV / Sn 140 kV, and 140 kV / 80 kV, on the contrast-to-noise ratio.

Fig. 9. Virtual monochromatic image had the highest CNR value, at (a) 68 keV under 80 kV / Sn 140 kV, (b) 72 keV under 100 kV / Sn 140 kV, and (c) 67 keV under 140 kV / 80 kV. obtained of (a) 1.571, (b) 1.717, and (c) 1.705, on CNR value.

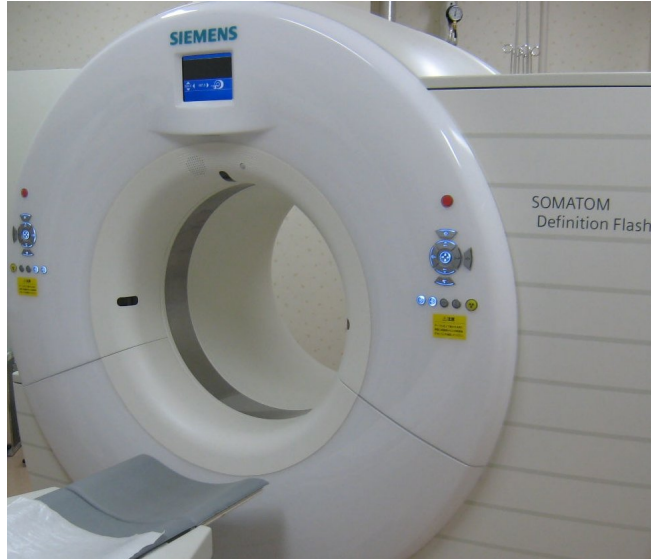

Fig. 1

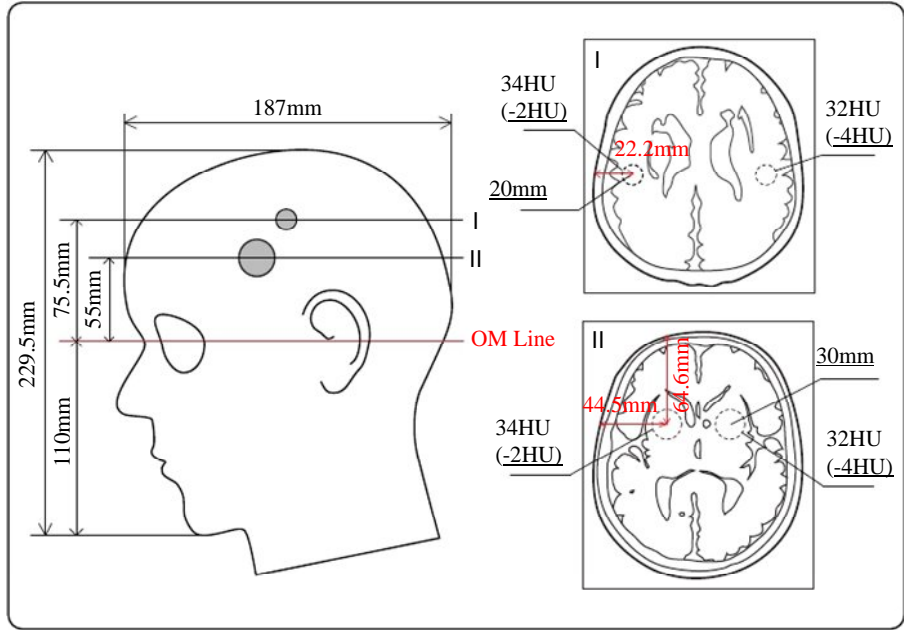

Fig. 2

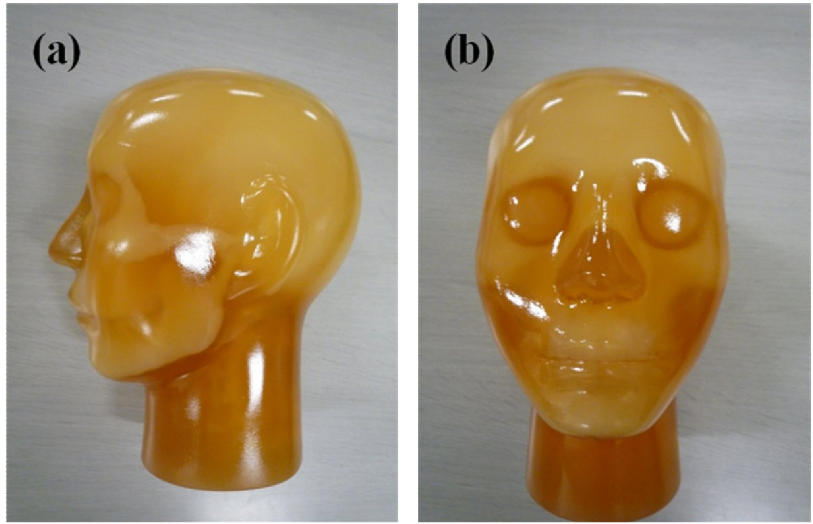

Fig. 3

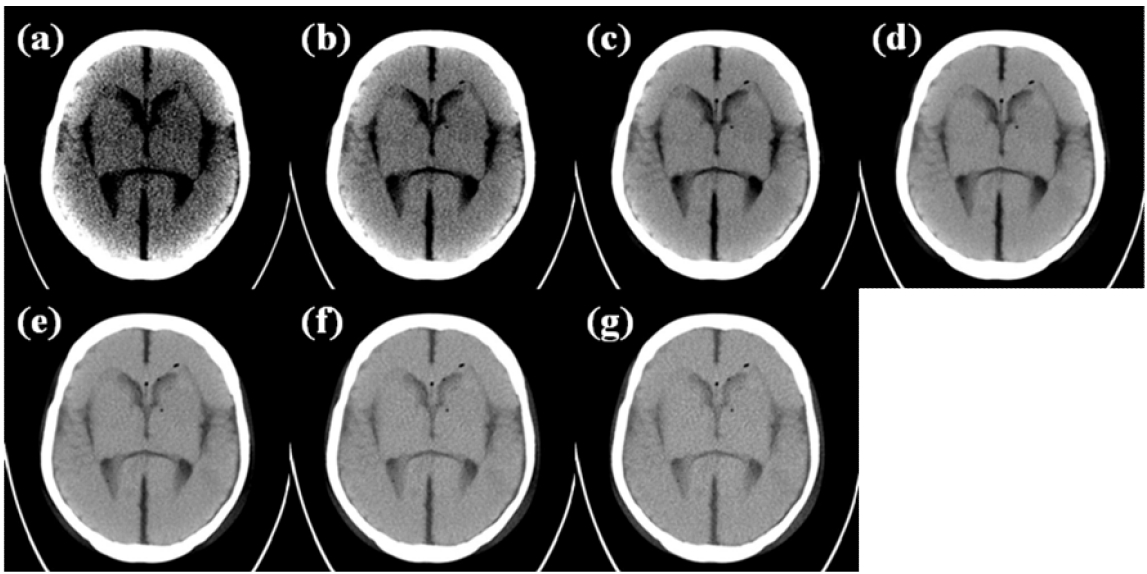

Fig. 4

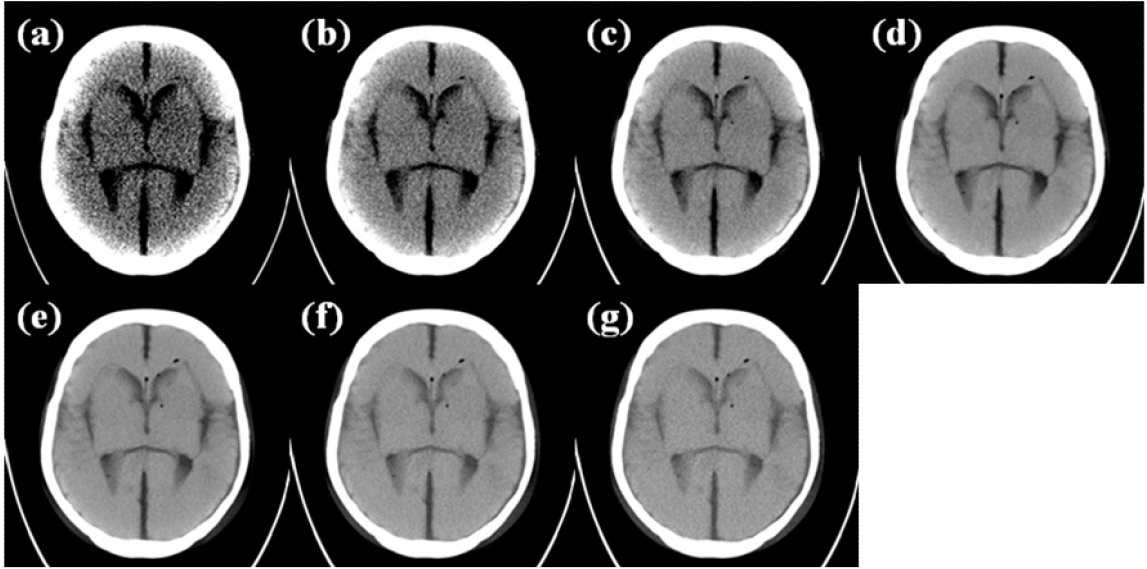

Fig. 5

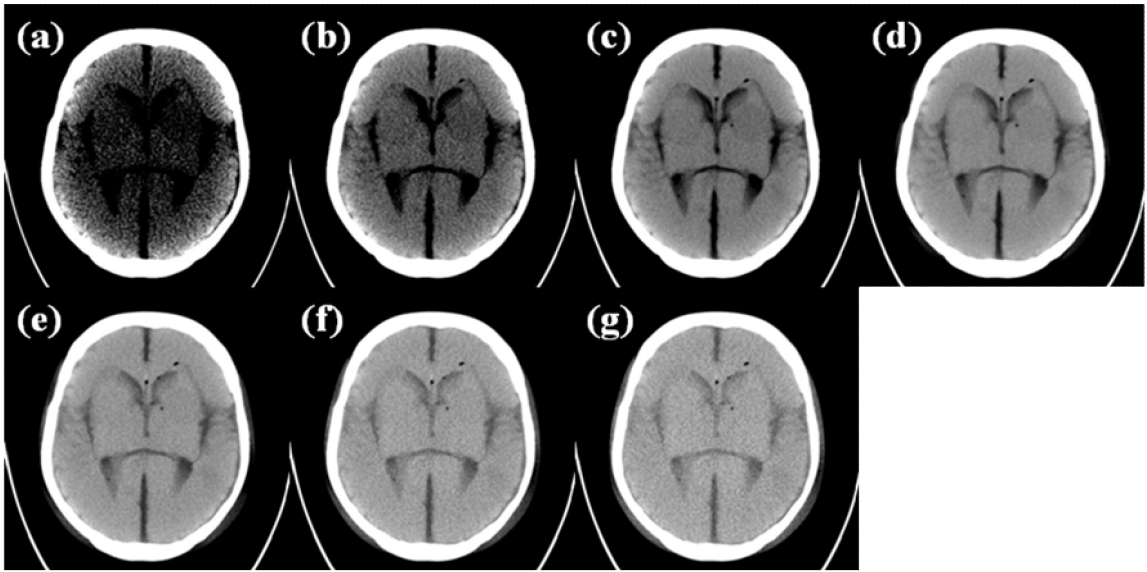

Fig. 6

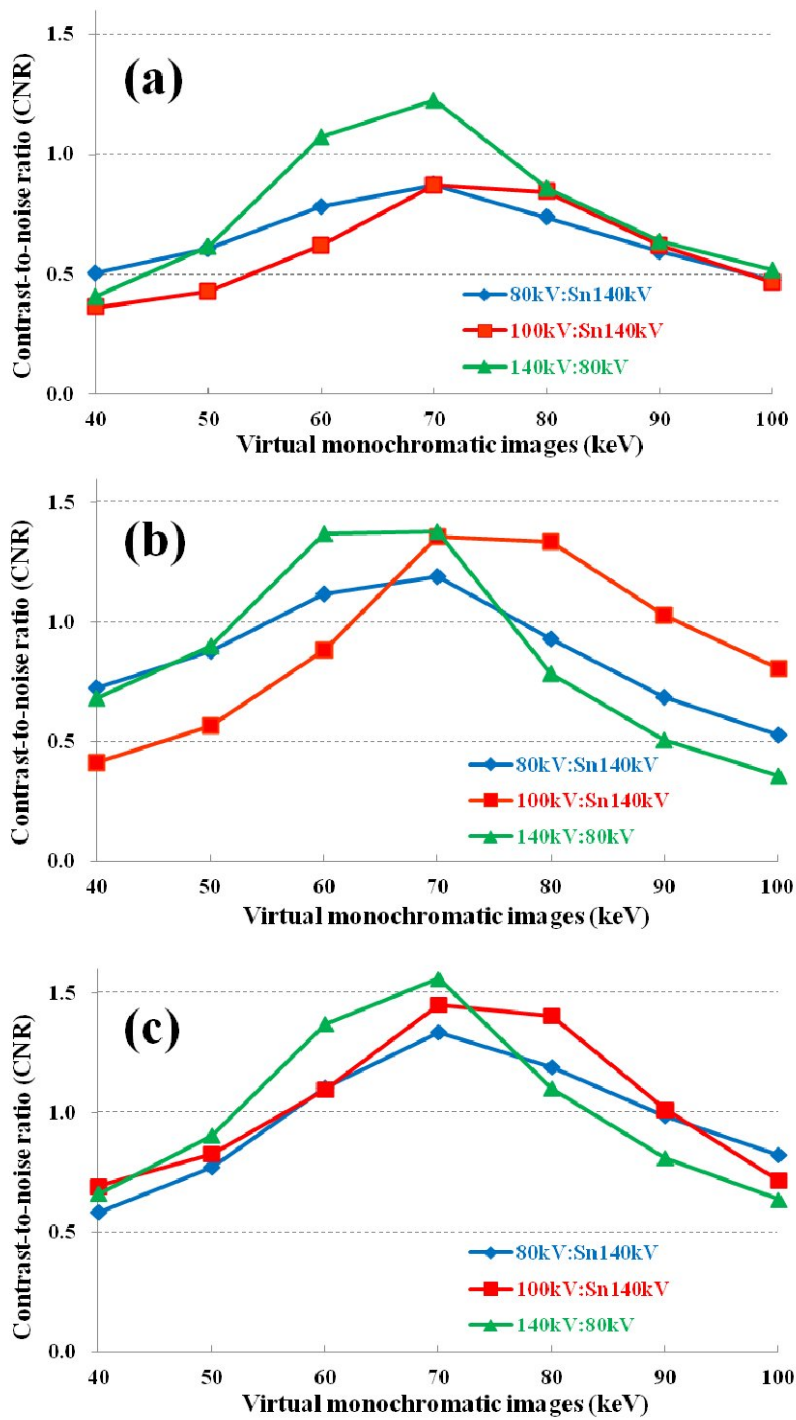

Fig. 7

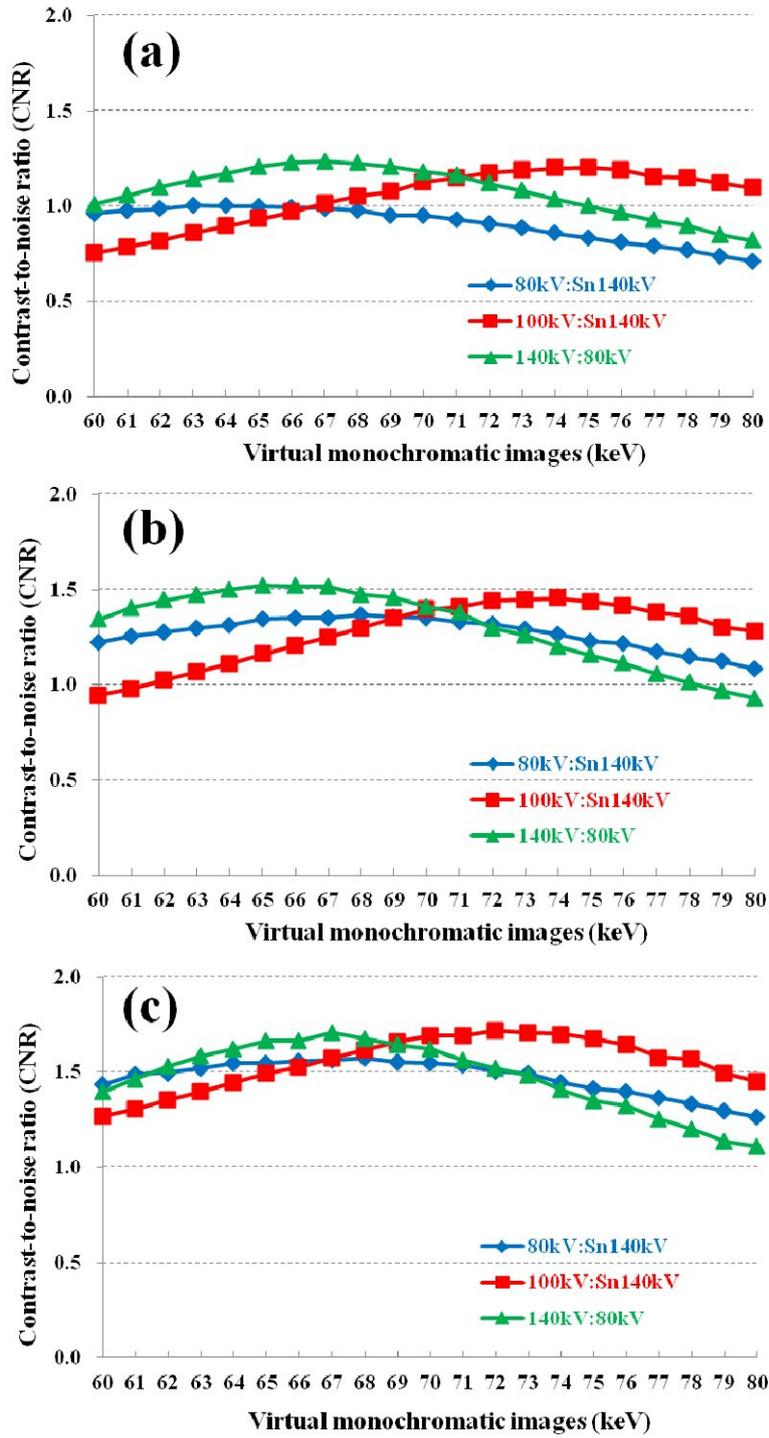

Fig. 8

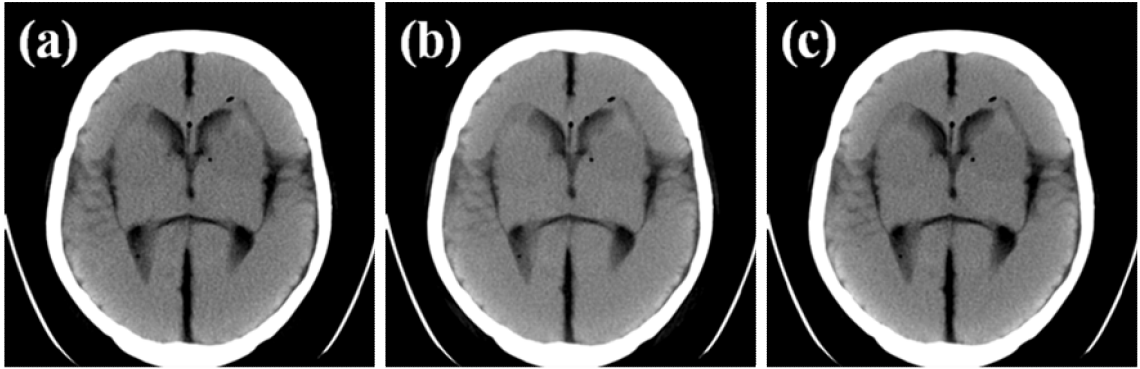

Fig. 9